\documentclass[showpacs,eqsecnum,aps]{revtex4}
\usepackage{amsmath}
\usepackage[dvips]{graphicx,color}

\begin{document}

\title{Subtractive renormalization of the NN scattering amplitude at leading
order in chiral effective theory}
\author{C.-J.~Yang, Ch.~Elster, and D.~R.~Phillips}
\affiliation{Institute of Nuclear and Particle Physics, and Department of Physics and
Astronomy, Ohio University,\\
Athens, OH 45701}
\email{cjyang, elster, phillips@phy.ohiou.edu}
\date{\today}

\begin{abstract}
The leading-order nucleon-nucleon (NN) potential derived from chiral
perturbation theory consists of one-pion exchange plus
short-distance contact interactions. We show that in the $^1$S$_0$ and
$^3$S$_1$--$^3$D$_1$ channels renormalization of the
Lippmann-Schwinger equation for this potential can be achieved by
performing one subtraction. This subtraction requires as its only
input knowledge of the NN scattering lengths. This procedure leads to
a set of integral equations for the partial-wave NN t-matrix which
give cutoff-independent results for the corresponding NN phase
shifts. This reformulation of the NN scattering equation offers
practical advantages, because only observable quantities appear in the
integral equation. The scattering equation may then be analytically
continued to negative energies, where information on bound-state
energies and wave functions can be extracted.
\end{abstract}

\pacs{12.39.Fe, 25.30.Bf, 21.45.+v }
\maketitle

\vspace{10mm} 




\section{Introduction}

\label{sec-intro}

Chiral perturbation theory
($\chi$PT)~\cite{Weinberg:1978kz,Gasser:1984gg,Bernard:1995dp}
provides a systematic technique by which nuclear forces can be
derived. As first pointed out by Weinberg~\cite{We90, We91}, the
nucleon-nucleon (NN) potential  can be expanded in powers of the
$\chi$PT expansion parameter $P \equiv
\frac{p,m_\pi}{\Lambda_\chi}$. The leading-order (LO=$P^0$) potential
is then the venerable one-pion exchange (OPE), supplemented by a
short-distance ``contact interaction'', which is a constant in
momentum space, or equivalently in co-ordinate space a
three-dimensional delta function.  Thus, the LO $\chi$PT potential is
singular, and when iterated in a Lippmann-Schwinger (LS) equation 
requires regularization and renormalization.

Weinberg's idea has since been extended to several orders in the
chiral expansion with NN potentials of $\chi$PT derived to NNLO in
Refs.~\cite{Or96,Ka97,Ep99} and to N$^3$LO~\cite{EM03,Ep05}. At
N$^3$LO a fit to NN data below $T_{\mathrm{lab}}=200$ MeV can be
achieved which is comparable in quality to that obtained in
`high-quality' NN potential models, provided cutoffs in the range
500--600 MeV are considered~\cite{EM03}. For this cutoff range the
phase shifts obtained from the N$^3$LO $\chi$PT potential have little
residual cutoff dependence at $T_{\mathrm{lab}} \leq 200$ MeV.

However, cutoff independence over a wider range is desirable. In
field-theoretic terms the issue of whether this can be achieved or not
speaks to whether a single momentum-independent contact interaction is
sufficient to renormalize the singular OPE potential~\cite{Le94,Le97}.
Beane, Bedaque, Savage, and van Kolck imposed a co-ordinate space
regulator on the LO $\chi$PT potential, and adjusted the two free
parameters that multiply the contact interactions so as to reproduce
both S-wave NN scattering lengths. They then showed that the resulting
phase shifts are independent of the regulator scale (up to small
higher-order corrections) when it is varied over an order of
magnitude~\cite{Be02}.  (Ref.~\cite{Be02} also showed that
renormalization of the $m_\pi $-dependence of the nuclear force
requires the contact interaction to be a function of $m_\pi$, but such
$m_\pi$-dependence will not concern us here.) This cutoff-independence
of NN S-wave phase shifts resulting from the LO potential derived by
Weinberg has since been confirmed by Ruiz Arriola and Pavon
Valderrama~\cite{PVRA04}, and---using a momentum-space cutoff---by
Nogga, Timmermans and van Kolck~\cite{NTvK05}. In addition,
Ref.~\cite{NTvK05} examined the behavior of this LO potential in
higher partial waves and argued that channels with $l \geq 1$ in which
the OPE potential is attractive also need additional contact
interactions if they are to be properly renormalized, a finding which
was confirmed in Ref.~\cite{PVRA06A}. This analysis agrees with an
examination of the renormalization-group scaling of short-distance
operators in the presence of one-pion exchange~\cite{Birse2006}.  The
usefulness of the results of Refs.~\cite{NTvK05,PVRA06A,Birse2006} has,
however, been disputed in Ref.~\cite{EM06}. For a review of the
situation see Ref.~\cite{Ha06}.

However, the NN partial waves on which this ongoing discussion centers
are not the focus of this study. Instead, we will re-examine the
renormalization of the leading-order $\chi$PT potential in the
$^1$S$_0$ and $^3$S$_1$-$^3$D$_1$ channels. We will develop a strategy
that replaces the divergent integrals and unphysical contact
interactions that appear in the LS equation for this potential with
finite integrals and physical quantities.  In particular, we will show
that knowledge of the zero-energy on-shell NN scattering amplitude
allows us to turn an ill-behaved LS equation into a set of equations
that give cutoff-independent results which are equivalent to the ones
obtained in Ref.~\cite{NTvK05}.

Such a conversion of divergent loop integrals and potentially infinite
Lagrangian couplings into a sum of experimental quantities and
convergent integrals is a well-known strategy in perturbation
theory. In a renormalizable theory it was shown to be possible to
pursue this strategy to all orders in perturbation theory by
Bogoluibov, Parasuik, Hepp, and Zimmerman~\cite{IZ}. Hereafter we refer to this
approach as ``subtractive'' renormalization. We show that such
subtractive renormalization can be applied to the LS equation for the
LO $\chi$PT potential. This is a non-trivial result, as while contact
interactions have simple integral representations in perturbation
theory the contact terms appearing in the LO $\chi$PT potential do
not. In consequence the development of subtractive renormalization in
a non-perturbative context is, to our knowledge, only a recent
advance, having been originally developed for the ``pionless EFT'' in
Refs.~\cite{HM99,AP03}.

An earlier attempt to develop subtractive renormalization for the NN
interaction of $\chi$PT was made in Ref.~\cite{Ti99}, and extended in
Ref.~\cite{Ti05}.  
There a large negative energy, $E_{sub}=-\frac{\mu^{2}}{M}$, is used as
subtraction point while solving the LS equation at a fixed c.m. energy in the NN
system.
The assumption that the Born approximation holds at $E_{sub}$, provided the
parameter $\mu$ is large enough, then determines the behavior of the NN amplitude
at that energy. However, this assumption is not generally
valid for a singular potential. Frederico \textit{et
  al.}  therefore effectively enforce a particular off-shell
dependence of the t-matrix at $E_{sub}$. We will show that such
an assumption is not necessary.  Instead, subtractions first suggested
by Hammer and Mehen~\cite{HM99} can be used at $E=0$ to obtain the
half-shell NN $t$-matrix $t(p,0;E=0)$ from the original LS equation
and the NN scattering length. From this, NN phase shifts at any energy
can be be obtained assuming only Hermiticity of the underlying
potential.  The latter technique was developed and employed by Afnan
and Phillips~\cite{AP03} in the context of the low-energy three-body
problem, but was not yet used in a two-body context.

The paper is structured as follows. In Section~\ref{sec-LONN} we
review the LO $\chi$PT $NN$ potential and solve the LS equation for
S-waves employing a momentum cutoff. We demonstrate the well-known
result that S-wave phase shifts are cutoff dependent unless the
contact interaction is adjusted so as to reproduce some low-energy observable. We
reproduce the results of Ref.~\cite{NTvK05} for cutoffs in the range
0.5 to 4 GeV. In Sec.~\ref{sec-subtract} we perform subtractions on
the LS equation such that reference to the contact interaction
disappears and demonstrate that this procedure results in
cutoff-independent predictions for NN phase shifts.  Our use of
subtractive renormalization allows us to consider cutoffs that are
much larger than those used in Ref.~\cite{NTvK05}, and in
Sec.~\ref{sec-subtract} we consider $\Lambda$ as high as 50 GeV in
order to show what is possible with our new technique. We also explain
the differences between our approach and that of Ref.~\cite{Ti99}.  In
Sec.~\ref{sec-bound} we analytically continue the subtracted equation
to negative energies and obtain bound-state energies and wave
functions, thereby showing that our techniques are equally suitable 
for obtaining the NN bound state energy and wave function.  We
summarize and outline possible extensions of this work in
Sec.~\ref{sec-conclusion}.



\section{The leading-order NN potential in chiral perturbation theory}

\label{sec-LONN}

As first pointed out by Weinberg in Refs.~\cite{We90,We91}, in
$\chi$PT the operators of lowest chiral dimension which can contribute
to the NN potential are:
\begin{equation}
\langle \mathbf{p}^{\prime }|V|\mathbf{p}\rangle =-\tau _{1}\cdot \tau _{2}
\frac{g_{A}^{2}}{4f_{\pi }^{2}}\frac{\sigma _{1}\cdot (\mathbf{p}^{\prime }-
\mathbf{p})\sigma _{2}\cdot (\mathbf{p}^{\prime }-\mathbf{p})}{(\mathbf{p}
^{\prime }-\mathbf{p})^{2}+m_{\pi }^{2}}+ \pi [(C_{T}+3C_{S})+\tau
_{1}\cdot \tau _{2}(C_{S}-C_{T})]  \label{eq:opep}
\end{equation}
where $\tau _{1},\tau _{2}$ are the isospin and  $\sigma _{1},\sigma _{2}$
the spin operators. The couplings constants are given by
 $g_{A\text{ }}=1.25$ and $f_{\pi }=93$ MeV, the c.m. momentum by $\mathbf{p}$. 
The low-energy constants $C_S$ and $C_T$ represent the effects of
degrees of freedom with masses of order $\Lambda_{\chi}$. Their
value is not determined by chiral symmetry, and they must be fitted to
experimental or lattice~\cite{NPLQCD} data.
Weinberg then advocated inserting
this potential into the non-relativistic Lippmann-Schwinger equation, 
which is sufficient because we are considering a low-energy expansion
of $V$:
\begin{equation}
\langle \mathbf{p}^{\prime }|t(E)|\mathbf{p}\rangle =\langle \mathbf{p}
^{\prime }|V|\mathbf{p}\rangle +\int \frac{d^{3}p^{\prime \prime }}{(2\pi
)^{3}}\langle \mathbf{p}^{\prime }|V|\mathbf{p}^{\prime \prime }\rangle 
\frac{1}{E+i\varepsilon -\frac{p^{\prime \prime }{}^{2}}{M}}\langle \mathbf{p
}^{\prime \prime }|t(E)|\mathbf{p}\rangle   \label{eq:LSE}
\end{equation}
(with $\varepsilon$ a positive infinitesimal and $M$ the nucleon mass).
This resums the effects of infra-red enhancement due to the
presence of heavy (c.f. $m_{\pi }$) particles in the intermediate NN
state. In a partial-wave expansion
Eq.~(\ref{eq:LSE}) reads
\begin{equation}
t_{l^{\prime }l}(p^{\prime },p;E)=v_{l^{\prime }l}(p^{\prime
},p)+\sum_{l^{\prime \prime }}\frac{2}{\pi }M\int_{0}^{\infty }\frac{
dp^{\prime \prime }\;p^{\prime \prime }{}^{2}\;v_{ll^{\prime \prime
}}(p^{\prime },p^{\prime \prime })\;t_{l^{\prime \prime }l}(p^{\prime \prime
},p,E)}{p_{0}^{2}+i\varepsilon -p^{\prime \prime }{}^{2}},
\label{eq:2.3}
\end{equation}
where $p_{0}^{2}/M=E$ is the c.m. energy, and $l$ denotes the
angular-momentum quantum number. For the singlet state we have
$l=l^{\prime }=l^{\prime \prime}=0$, for the triplet state these
values can be 0 or 2. The explicit expressions for the partial wave
projections of OPE are given in Ref.~\cite{Ti99}. For the
convenience of the reader we repeat them here. For the $^{1}$S$_{0}$
state one obtains $($with $C_{T},C_{S}=0)$
\begin{equation}
v_{00}^{s}(p^{\prime },p)=\frac{g_{A}^{2}}{32\pi f_{\pi }^{2}}\left(
2-\int_{-1}^{1}dx\frac{m_{\pi }^{2}}{p^{\prime }{}^{2}+p^{2}-2p^{\prime
}px+m_{\pi }^{2}}\right) .  \label{eq:OPEs}
\end{equation}
The angular-momentum projected matrix elements for the $^{3}$S$_{1}$-$^{3}$D$
_{1}$ channels ($l$,$l^{\prime }$ = 0,2) are given by 
\begin{eqnarray}
v_{00}^{t}(p^{\prime },p) &=&\frac{g_{A}^{2}}{32\pi f_{\pi }^{2}}\left(
2-\int_{-1}^{1}dx\frac{m_{\pi }^{2}}{p^{\prime }{}^{2}+p^{2}-2p^{\prime
}px+m_{\pi }^{2}}\right)  \notag \\
v_{20}^{t}(p^{\prime },p) &=&\frac{g_{A}^{2}\sqrt{2}}{16\pi f_{\pi }^{2}}
\left( \int_{-1}^{1}dx\frac{p^{\prime }{}^{2}-2p^{\prime }px+p^{2}(\frac{3}{2
}x^{2}-\frac{1}{2})}{p^{\prime }{}^{2}+p^{2}-2p^{\prime }px+m_{\pi }^{2}}
\right)  \notag \\
v_{22}^{t}(p^{\prime },p) &=&\frac{g_{A}^{2}}{32\pi f_{\pi }^{2}}\left(
\int_{-1}^{1}dx\frac{2pp^{\prime }x-(p^{2}+p^{\prime }{}^{2})(\frac{3}{2}
x^{2}-\frac{1}{2})}{p^{\prime }{}^{2}+p^{2}-2p^{\prime }px+m_{\pi }^{2}}
\right).  \label{eq:OPEt}
\end{eqnarray}
Inspection of the limits for $p,p^{\prime }\rightarrow \infty $ for $
v_{00}(p^{\prime },p)$ in Eq.~(\ref{eq:OPEs}) gives \cite{Ti99} 
\begin{equation}
\lim_{p\text{ or }p^{\prime }\rightarrow \infty }v_{00}^{s}(p^{\prime },p)=
\frac{g_{A}^{2}}{16\pi f_{\pi }^{2}}.  \label{eq:2.6}
\end{equation}
For the triplet channels the following limits give a finite value \cite{Ti99}
\begin{eqnarray}
\lim_{p\;\mathrm{or}\;p^{\prime }\rightarrow \infty
}v_{00}^{t}(p^{\prime},p) &=&\frac{g_{A}^{2}}{16\pi f_{\pi }^{2}},
\notag \\ \lim_{p^{\prime }\rightarrow \infty }v_{20}^{t}(p^{\prime
},p) &=&\frac{ g_{A}^{2}\sqrt{2}}{8\pi f_{\pi }^{2}}, \label{eq:2.7}
\end{eqnarray}
while $\lim_{p\rightarrow \infty }v_{20}^{t}(p^{\prime
},p)=\lim_{p,p^{\prime }\rightarrow \infty }v_{22}^{t}(p^{\prime
},p)=0$.  Thus, the integrals of Eq.~(\ref{eq:2.3}) diverge. The
ultra-violet behavior of the integral equation
can be regulated by introducing a momentum cutoff $\Lambda $
as the upper bound of the integration. Meanwhile, the infra-red
singularity at $p=p_0$ in Eq.~(\ref{eq:2.3}) is separated into a
principal-value part and a delta-function part, with the principal-value
singularity treated by standard subtraction techniques
\cite{hafteltabakin}. 

\begin{figure}(10,6)
\begin{center}
\includegraphics[width=9cm]{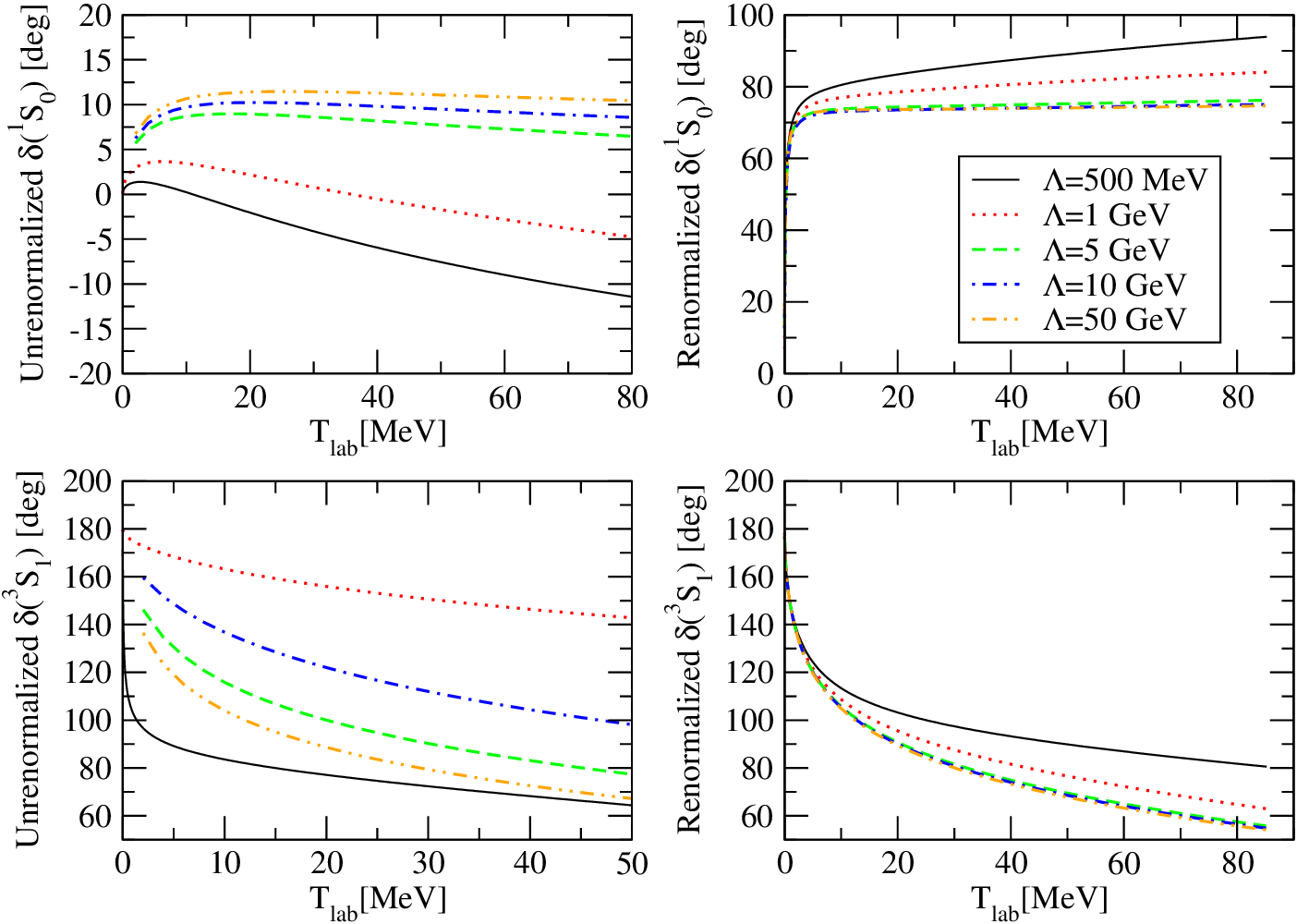}
\end{center}
\caption{(Color online) The NN phase shift $\protect\delta (^{1}$S$_{0})$ and 
$\protect\delta (^{3}$S$_{1})$ as function of
the laboratory kinetic energy $T_{\mathrm{lab}}\leq 80$ MeV for cutoff
parameters $\Lambda $ ranging from 0.5--50~GeV. The upper left panel shows the
phase shift with $C_S=0$, while the upper right panel shows the value with
$C_S$ adjusted at each value of $\Lambda$ to reproduce the
experimental value of $a_s$. The bottom
left panel shows the phase shift with $C_T=0$, while the bottom right
panel shows the value with $C_T$ adjusted at each value of $\Lambda$
to reproduce the experimental value of $a_t$. Note that the phase
shift in both lower panels is given modulo $\pi$ only, since multiple
deeply bound states are present.}
\label{fig-fig1}
\end{figure}

Once these two steps are made the integral equation can be solved by
standard Gauss-Legendre quadrature.  The results of our numerical
calculations for the $^{1}$S$_{0}$ and $^{3}$S$_{1}$-$^{3}$D$_{1}$
phase shifts are shown on the left-hand side of Figs.~\ref{fig-fig1}
. We confirm that if $C_S$($C_T$) takes a fixed
value (e.g. zero), the resulting ${}^1$S$_0$ (${}^3$S$_1$) phase shift
has sizable cutoff variation---of order $\delta_{^{1}S_{0}}$ ($\delta
_{^{3}S_{1}}$) itself---even for small projectile laboratory energies
$T_{\mathrm{lab}}$.

One way to remedy the situation is to renormalize each of the S-wave
potentials by taking advantage of the presence of the arbitrary
constants $C_{S} $ and $C_{T}$. In Ref.~\cite{NTvK05}
Eq.~(\ref{eq:LSE}) was solved with the relevant constant adjusted such
that for a given cutoff $\Lambda $ the S-wave phase shifts at 10 MeV
were reproduced. Here we do something similar, except that we use the
$NN$ scattering lengths as input, adopting the values $a_{s}=-23.7$~fm
for the singlet case and $a_{t}=5.43$~fm for the triplet~\cite{St94}.

Once $C_S$ and $C_T$ are fitted in this way we can predict the phase
shift at an arbitrary energy, and the resulting ``renormalized'' phase
shifts are plotted in in the right-hand panels of Figs.~\ref{fig-fig1}
 as a function of the laboratory kinetic energy. The
phase shifts are now approximately cutoff independent, with the
dependence on $\Lambda$ relegated to effects $\sim 1/\Lambda$ in the
effective ranges, $r_s$ and $r_t$ (see Table~\ref{table-3}). This
reproduces results of Beane, Bedaque, Savage and van Kolck and
others~\cite{Be02,NTvK05}.

\begin{figure}[tbp]
\begin{center}
\includegraphics[width=7.7cm]{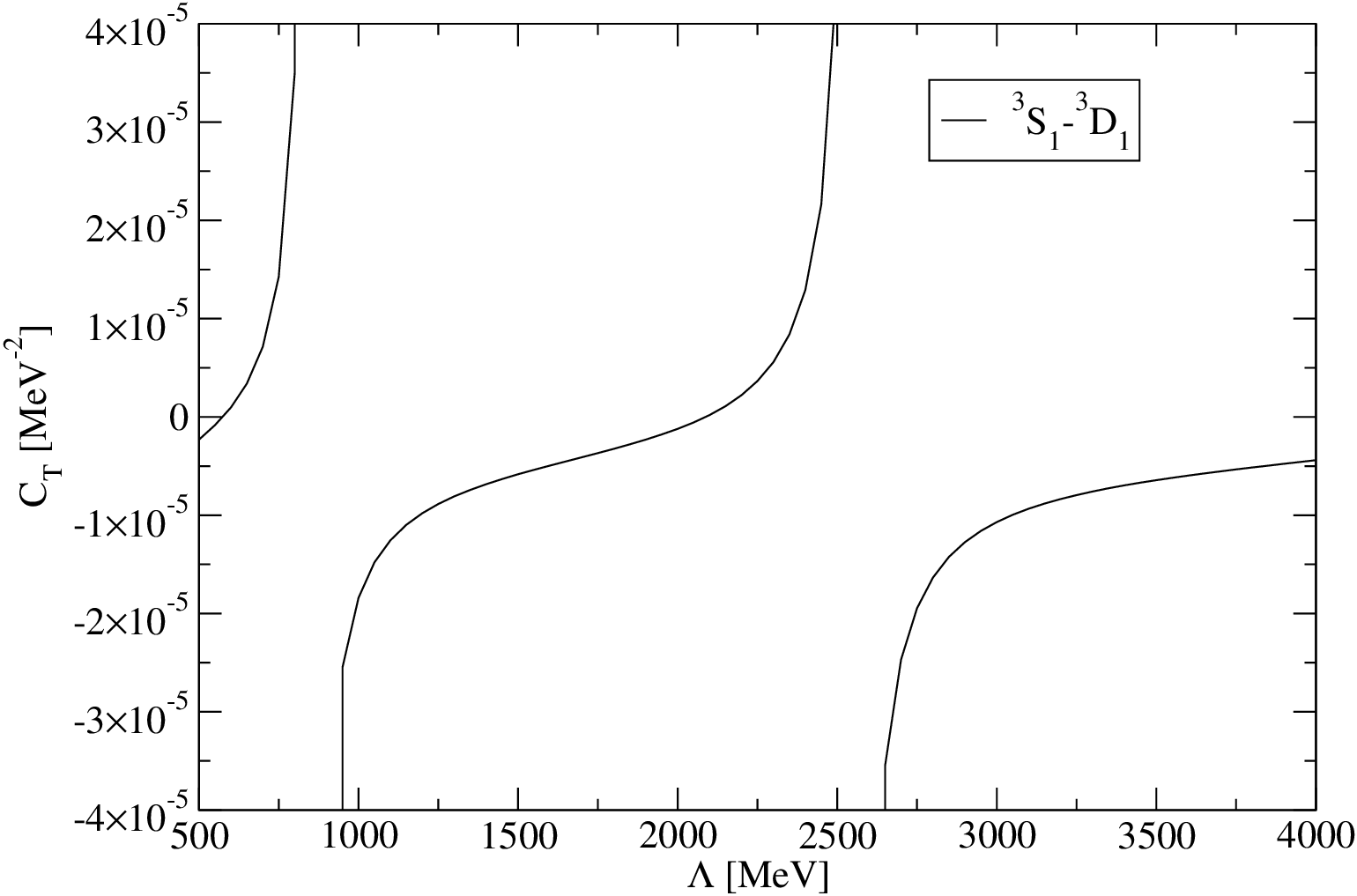}
\end{center}
\caption{The renormalization constant $C_T$  for the triplet channel  as a
function of the cutoff $\Lambda$, where $\Lambda$ ranges from 0.5--4~GeV. }
\label{fig-fig12}
\end{figure}

The renormalization constants $C_{S}$ and $C_{T}$ that were used to
generate these results are given in Table~\ref{table-1} together with
the corresponding cutoff $\Lambda$.  While $C_{S}$ exhibits a 
monotonic behavior as function of $\Lambda$, the evolution of $C_T$ with
$\Lambda$ is particularly interesting, and is shown in
Fig.~\ref{fig-fig12}. This confirms the result shown in Fig.~2 of
Ref.~\cite{NTvK05}, and shows that $C_T(\Lambda)$ diverges at specific
values of $\Lambda$.  The origin of this behavior is the appearance of
additional bound states in the $^3$S$_1$-$^3$D$_1$ channel as the cutoff is
increased. At the specific values of $\Lambda$ where one more bound
state appears, Eq.~(\ref{eq:LSE}) cannot be solved by fitting $C_T(\Lambda)$
to reproduce $a_t$.

But even when the constants $C_{T(S)}$ are finite, fitting them
in order to reproduce the scattering length can be very delicate.
For example, at $\Lambda=1$ GeV $C_S$ must be tuned to 1
part in $10^4$ in order to get $a_s$ accurate to 1\%, while for
$\Lambda=10$ GeV it must be tuned to 1 part in 10$^6$ to achieve this
accuracy. At arbitrarily large cutoffs it therefore becomes numerically
impossible to fit $C_S$ and $C_T$.

\section{Subtracted Integral Equations}

\label{sec-subtract}

An alternative to fitting the contact interactions $C_S$ and $C_T$ so
as to reproduce the ${}^1$S$_0$ and ${}^3$S$_1$ scattering lengths is
to perform subtractive renormalization on the LS equations for the
potential (\ref{eq:opep}). In this section we explain the 
subtractive renormalization procedure. Note that in contrast
to the approach advocated in Ref.~\cite{Ti99}, our subtraction
procedure does not require assumptions about the behavior of $t$ 
 anywhere.

Let us first concentrate 
on the c.m. energy $E=0$ in the NN system.
The definition of the scattering lengths $a_{s(t)}$
in the limit $E \equiv p_{0}^{2}/M\rightarrow 0$ gives 
in the singlet (triplet) states
\begin{equation}
t_{00}^{s(t)}(0,0,E=0)=\frac{f_{00}^{s(t)}(0)}{-Mp_{0}}=\frac{a_{s(t)}}{M},
\label{eq:3.1}
\end{equation}
where $f_{00}^{s(t)}(E)$ represents the scattering amplitude in the singlet (s)
or triplet (t) channel. The half-shell and on-shell LS equations for this energy
read
\begin{eqnarray}
t(p,0;0) &=&v(p,0)+C +\frac{2}{\pi }M\int_{0}^{\Lambda} dp^{\prime}
\;p^{\prime }{}^{2}\left( \frac{v(p,p^{\prime })+ C}{-p^{\prime
}{}^{2}}\right) \;t(p^{\prime },0;0)  \label{eq:3.2} \\
t(0,0;0) &=&v(0,0)+C +\frac{2}{\pi }M\int_{0}^{\Lambda }dp^{\prime}
\;p^{\prime }{}^{2}\left( \frac{v(0,p^{\prime })+C}{-p^{\prime
}{}^{2}}\right) \;t(p^{\prime },0;0),  \label{eq:3.3}
\end{eqnarray}
where we dropped
the channel  indices since our procedure is valid for the singlet as well as
triplet LS equations.
 Subtracting Eq.~(\ref{eq:3.3}) from
Eq.~(\ref{eq:3.2}) and using (\ref{eq:3.1})
leads to 
\begin{equation}
t_{00}^s(p,0;0)=v_{00}^s(p,0)+\frac{a_s}{M}+\frac{2}{\pi }M\;\int_{0}^{\Lambda }dp^{\prime
}\;p^{\prime }{}^{2}\;\left( \frac{v_{00}^s(p,p^{\prime
  })-v_{00}^s(0,p^{\prime })}
{-p^{\prime }{}^{2}}\right) \;t_{00}^s(p^{\prime },0;0)  \label{eq:3.4}
\end{equation}
in the singlet channel, while for the triplet channel the equation
corresponding to Eq.~(\ref{eq:3.4}) is
\begin{equation}
t^t_{l^{\prime }l}(p^{\prime },0;0)-\sum_{l^{\prime \prime }}\frac{2}{\pi }
M\int_{0}^{\Lambda }dp^{\prime \prime }p^{\prime \prime 2}\left( \frac{
v^t_{l^{\prime }l^{\prime \prime }}(p^{\prime },p^{\prime \prime
})-v^t_{l^{\prime }l^{\prime \prime }}(0,p^{\prime \prime })}{-p^{\prime
\prime }{}^{2}}\right) t^t_{l^{\prime \prime }l}(p^{\prime \prime
},0;0)=v^t_{l^{\prime }l}(p^{\prime},0)- v^t_{l^\prime l}(0,0)
+\delta _{l^{\prime }0}\delta _{l0}
\frac{a_{t}}{M},  \label{eq:3.5}
\end{equation}
where $l,l^{\prime }=0,2$. Note that since $v_{20}(0,p;0)=v_{02}(p,0;0)=0$, and $%
v_{22}(0,p;0)=v_{22}(p,0;0)=0$, the corresponding t-matrix elements $t_{02}$
and $t_{22}$ must also be zero. 

Equations akin to Eqs.~(\ref{eq:3.4}) and (\ref{eq:3.5}) were first
developed for the three-body problem in Ref.~\cite{HM99}.  They have the
advantage that the constants $C_S$ and $C_T$ cancel out in the
differences that form Eqs.~(\ref{eq:3.4}) and (\ref{eq:3.5}), 
so that these equations contain only observable quantities. They
can be solved using standard techniques. Our results shows the
relative difference between the two methods is less than 0.2\% for all
channels.

Having determined $t(p,0;0)$ we set about obtaining half-shell t-matrix
elements $t(p,p^{\prime };0)$, $p^{\prime }$ being an arbitrary momentum, by
carrying out a further subtraction. First we note that:
\begin{eqnarray}
t(p,p^{\prime };0)&=&v(p,p^{\prime })+ C +\frac{2}{\pi}
M\;\int_{0}^{\Lambda} dp^{\prime \prime} \;p^{\prime \prime}{}^{2}\;\left( 
\frac{v(p,p^{\prime \prime }) + C}{-p^{\prime \prime }{}^{2}}\right)
t(p^{\prime \prime },p^{\prime };0)  \label{eq:3.6} \\
t(0,p^{\prime };0)&=&v(0,p^{\prime })+ C +\frac{2}{\pi }
M\;\int_{0}^{\Lambda }dp^{\prime \prime }\;p^{\prime \prime }{}^{2}\;
\left( \frac{v(0,p^{\prime \prime })+ C}{-p^{\prime \prime }{}^{2}}\right)
t(p^{\prime \prime },p^{\prime };0).  \label{eq:3.7}
\end{eqnarray}
Subtracting Eq.~(\ref{eq:3.7}) from Eq.~(\ref{eq:3.6}) leads to 
\begin{equation}
t(p,p^{\prime };0)-t(0,p^{\prime };0)=v(p,p^{\prime })-v(0,p^{\prime })+
\frac{2}{\pi }M\;\int_{0}^{\Lambda }dp^{\prime \prime }\;p^{\prime \prime
}{}^{2}\;\left( \frac{v(p,p^{\prime \prime })-v(0,p^{\prime \prime })}{
-p^{\prime \prime }{}^{2}}\right) t(p^{\prime \prime },p^{\prime };0).
\label{eq:3.8}
\end{equation}

\begin{figure}[tbp]
\begin{center}
\includegraphics[width=10cm]{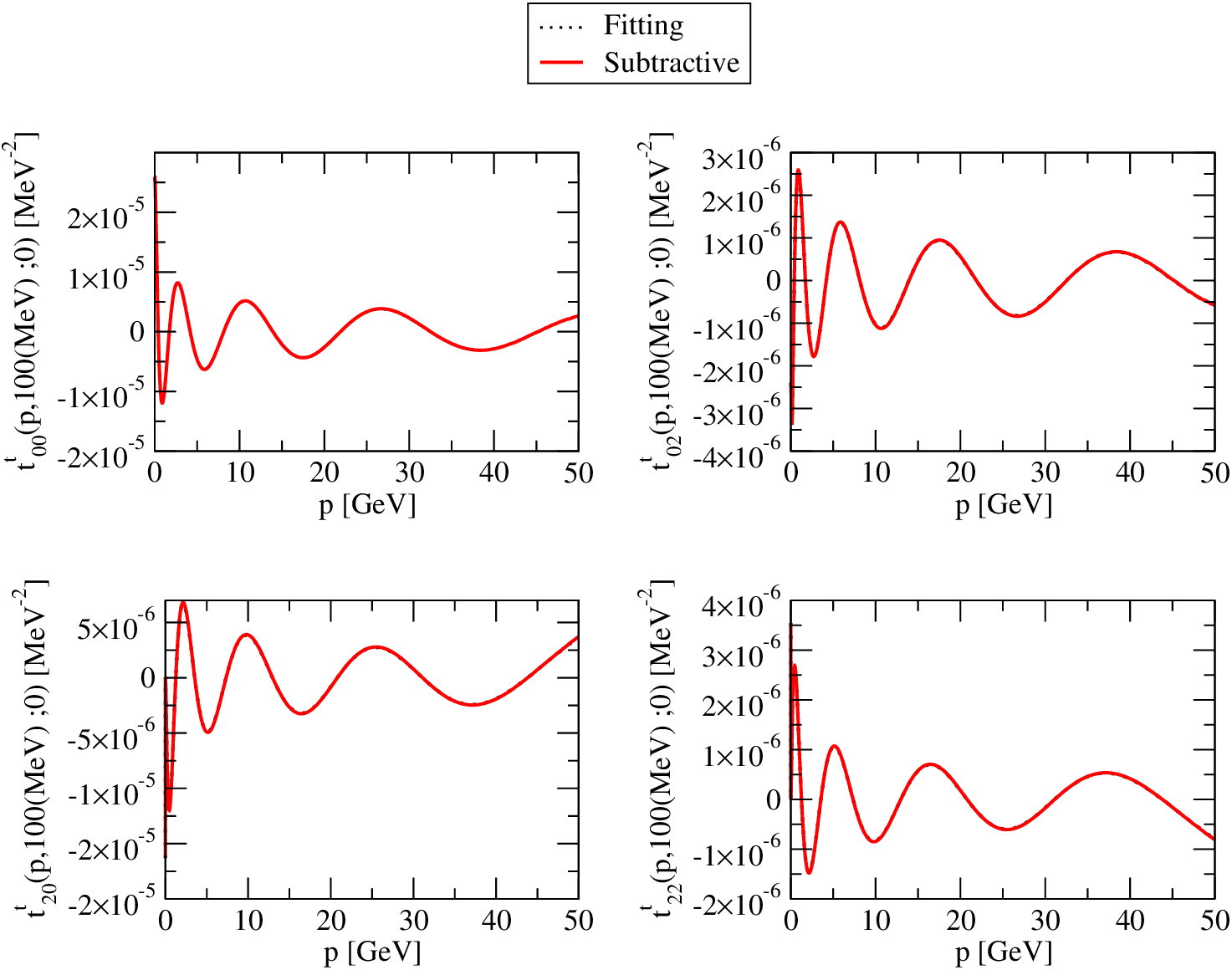}
\end{center}
\caption{(Color online) The comparison of the two renormalization  methods for zero energy off-shell
matrix element $t_{ll'}(p,p';0)$ in the triplet channel
as a function of the momentum $p$ for fixed $p'=$~100 MeV.}
\label{fig-fig4}
\end{figure}

Taking advantage of the property $t(p^{\prime },0;0)=t(0,p^{\prime
};0)^{T}$, we can use $t(p^{\prime },0;0)$ obtained from
Eq.~(\ref{eq:3.5}) to solve for $t(p,p^{\prime };0)$ in
Eq.~(\ref{eq:3.8}). The calculated results are displayed in
Fig.~\ref{fig-fig4} and
show good agreement (relative difference$\leq 2\%$) between this
subtractive method and the one described in the previous section.

\begin{figure}[tbp]
\begin{center}
\includegraphics[width=10cm]{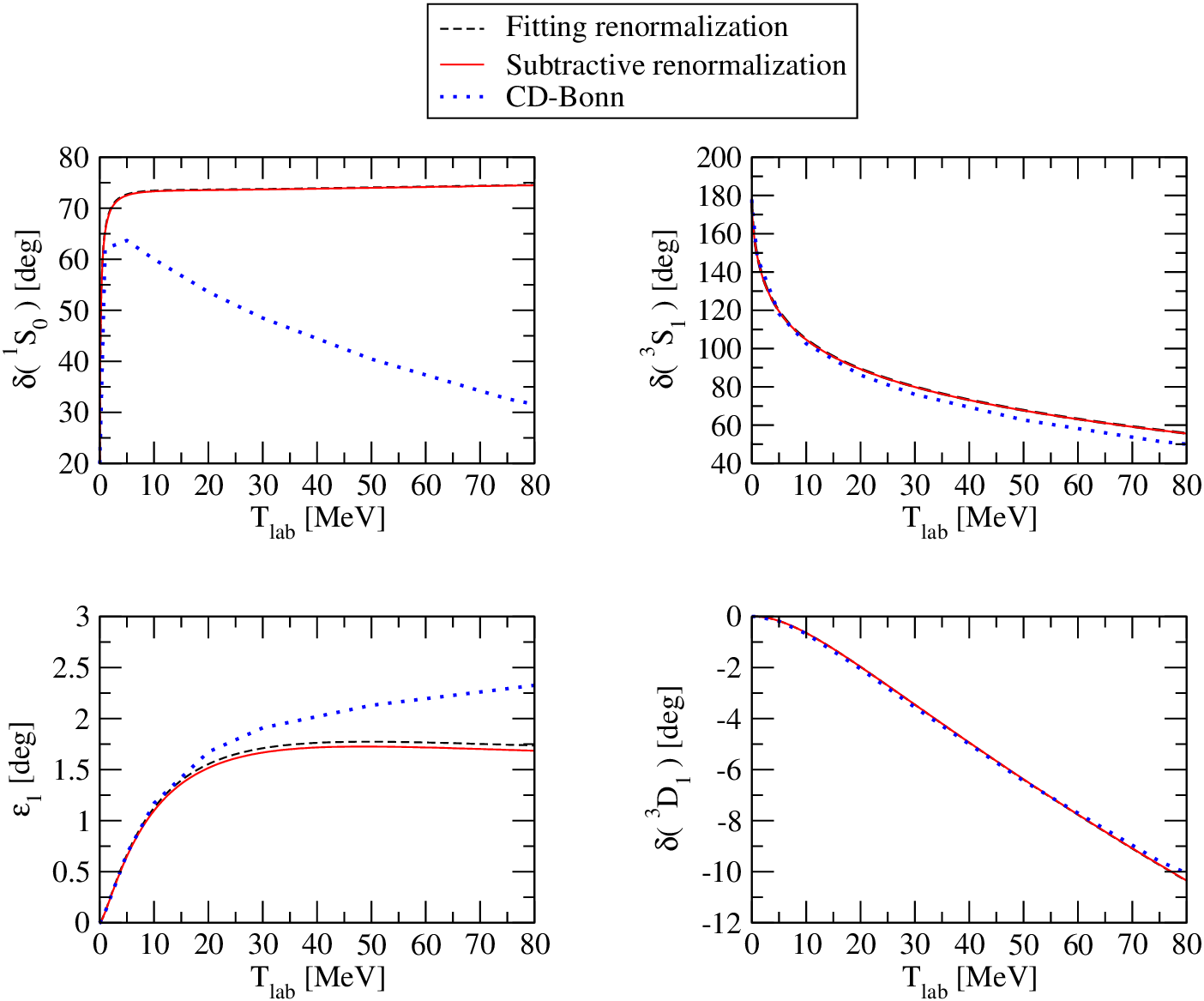}
\end{center}
\caption{(Color online) The comparison of two renormalization methods for the lowest
  NN singlet and triplet phase shifts and mixing parameter as a
  function of the laboratory kinetic energy $T_{\mathrm{lab}}\leq 80$
  MeV. Here $\Lambda=50$ GeV is used. 
The phase shifts obtained from the CD-Bonn potential are also
  shown (dashed lines).}
\label{fig-fig5}
\end{figure}

As a last step we need to obtain the general half-shell t-matrix
$t(p,p^{\prime};E)$ at arbitrary c.m. energies $E$ from the one at
zero energy that is given by Eq.~(\ref{eq:3.8}). We start again from
two LS equations,
\begin{eqnarray}
t(p,p^{\prime };0) &= &v(p,p^{\prime })+C +\frac{2}{\pi }
M\;\int_{0}^{\Lambda }dp^{\prime \prime }\;p^{\prime \prime }{}^{2}\;\left( 
\frac{v(p,p^{\prime \prime })+C}{-p^{\prime \prime }{}^{2}}\right)
\;t(p^{\prime },p^{\prime \prime };0).  \label{eq:3.9} \\
t(p,p^{\prime };E) &= &v(p,p^{\prime })+C +\frac{2}{\pi }
M\;\int_{0}^{\Lambda }dp^{\prime \prime }\;p^{\prime \prime }{}^{2}\;\left( 
\frac{v(p,p^{\prime \prime })+C}{p_{0}^{2}-p^{\prime \prime
}{}^{2}+i\epsilon }\right) \;t(p^{\prime },p^{\prime \prime };E),
\label{eq:3.10}
\end{eqnarray}
which, in order to express the idea more clearly, we simplify to
\begin{eqnarray}
t(0) &=&(v + C) + (v + C)g(0)t(0),  \label{eq:3.11} \\
t(E) &=&(v + C) +(v + C) g(E)t(E).  \label{eq:3.12}
\end{eqnarray}
The second equation, Eq.~(\ref{eq:3.12}), can be rewritten as 
\begin{equation}
t(E)(1+g(E) t(E))^{-1}=v + C,  \label{eq:3.13}
\end{equation}
whereas the first, Eq.~(\ref{eq:3.11}), can also be expressed as 
\begin{equation}
t(0)=(v + C) +t(0)g(0)(v+C).  \label{eq:3.14}
\end{equation}
Substituting Eq.~($\ref{eq:3.13}$) into Eq.~($\ref{eq:3.14}$) gives 
\begin{equation}
t(0)=\left[ 1+t(0)g(0)\right] \ t(E) \left[1+g(E)t(E)\right]^{-1},
\end{equation}
which after simplification yields
\begin{equation}
t(E)=t(0) + t(0)(g(E) - g(0))t(E).
\label{eq:3.15}
\end{equation}
We can now obtain $t(E)$ from $t(0)$ by solving the integral equation,
Eq.~(\ref{eq:3.15}). Then choosing $p=p_0$ with $E=p_0^{2}/M$ and
setting $p^{\prime}=p$ we have the on-shell t-matrix element
$t_l(p_0,p_0;E)$ at an arbitrary energy $E$.  The on-shell element is
related to the phase shift $\delta_l$ through the well-known relation
\begin{equation}
t_l(p_0,p_0;E)=-\frac{e^{i\delta_l }\sin{\delta_l} }{Mp_0}.
\end{equation}
The $^1$S$_0$ phase shifts calculated in this way is shown in the
upper left panel of Fig.~\ref{fig-fig5}, where they are compared to those
found by fitting $C_S$, as described in the previous section. The
agreement between the two methods is very good, which confirms the
ability of the subtraction to reproduce the results obtained by
fitting $C$.  The phase shifts produced by the CD-Bonn
potential~\cite{Machleidt:2000} are also shown in
Fig.~\ref{fig-fig5}. Since this potential reproduces the NN data in
this region with $\chi^2/{\rm d.o.f} \approx 1$ its $\delta_{{}^1S_0}$
can be regarded as a parameterization of experiments. But the CD-Bonn
values for $\delta_{^{1}S_{0}}$ are not well reproduced by the LO
$\chi$PT potential. In this channel the LO calculation obviously has
sizable higher-order corrections. In fact the description is
known to improve when additional terms in the chiral expansion of $V$
are included~\cite{Or96,Ep99,PVRA06B}.

Meanwhile the other three panels of Fig.~\ref{fig-fig5} show results 
for the triplet channel,  where we adopt the Stapp convention \cite{stapp}
for the phase shifts and the mixing parameter 
\begin{eqnarray}
\epsilon &=&\frac{1}{2}\arctan{\left[ \frac{-i(t_{02}^t+t_{20}^t)}{2\sqrt{%
t_{00}^tt_{22}^t}} \right] }  \notag \\
\delta (^{3}S_{1}) &=&\frac{1}{2}\arctan{\ \left[\frac{\Im[(t_{00}^t)/\cos
(2\epsilon )]}{\Re[(t_{00}^t)/\cos (2\epsilon )]}\right]}  \notag \\
\delta (^{3}D_{1}) &=&\frac{1}{2}\arctan \left[\frac{\Im[(t_{22}^t)/\cos
(2\epsilon )]}{\Re[(t_{22}^t)/\cos (2\epsilon )]}\right].
\end{eqnarray}
The agreement between the two methods employed to perform the LO
$\chi$PT calculation of these quantities is satisfactory, and for
$\delta_{{}^3S_1}$ and $\delta_{{}^3D_1}$ the agreement of the LO
calculation with the CD-Bonn phases is strikingly good.

The main numerical challenge of our subtractive renormalization scheme
originates in solving Eq.~(\ref{eq:3.8}) and Eq.~(\ref{eq:3.15}). 
Matrix elements with large $p'$ are small and in the subtraction scheme 
they are calculated from a combination of larger matrix elements.
  We now compare the efficiency of the subtraction and fitting methods.
We first note that due to the multi-step nature of subtraction, it does take about 3 times
 longer to run the code than it does for the fitting method with the same number of
mesh points. On the other hand, as Table~\ref{table-4} shows, the subtraction method converges much faster
than the fitting method with respect to the number of mesh points used in the calculation. With 20 mesh points
we already have 4 significant-figure accurancy for the subtraction method, while to reach this accurancy
it takes more than 200 mesh points for the fitting method. 
The subtraction method converges so much faster because in the fitting method $C_T$ affects only the part of the off-shell amplitude near $p=\Lambda$, and the oscillation phase
it sets there must be communicated down to $p=0$ to get a particular scattering length $a_{t}$. 
In the subtraction method the correct value of the scattering length is enforced in the equation itself, and 
does not have to be achieved by fine tuning the constant $C_T$ so as to obtain the correct behavior of the half-off-shell
amplitude for $p$ of order $\Lambda$.

\begin{figure}[tbp]
\begin{center}
\includegraphics[width=10cm]{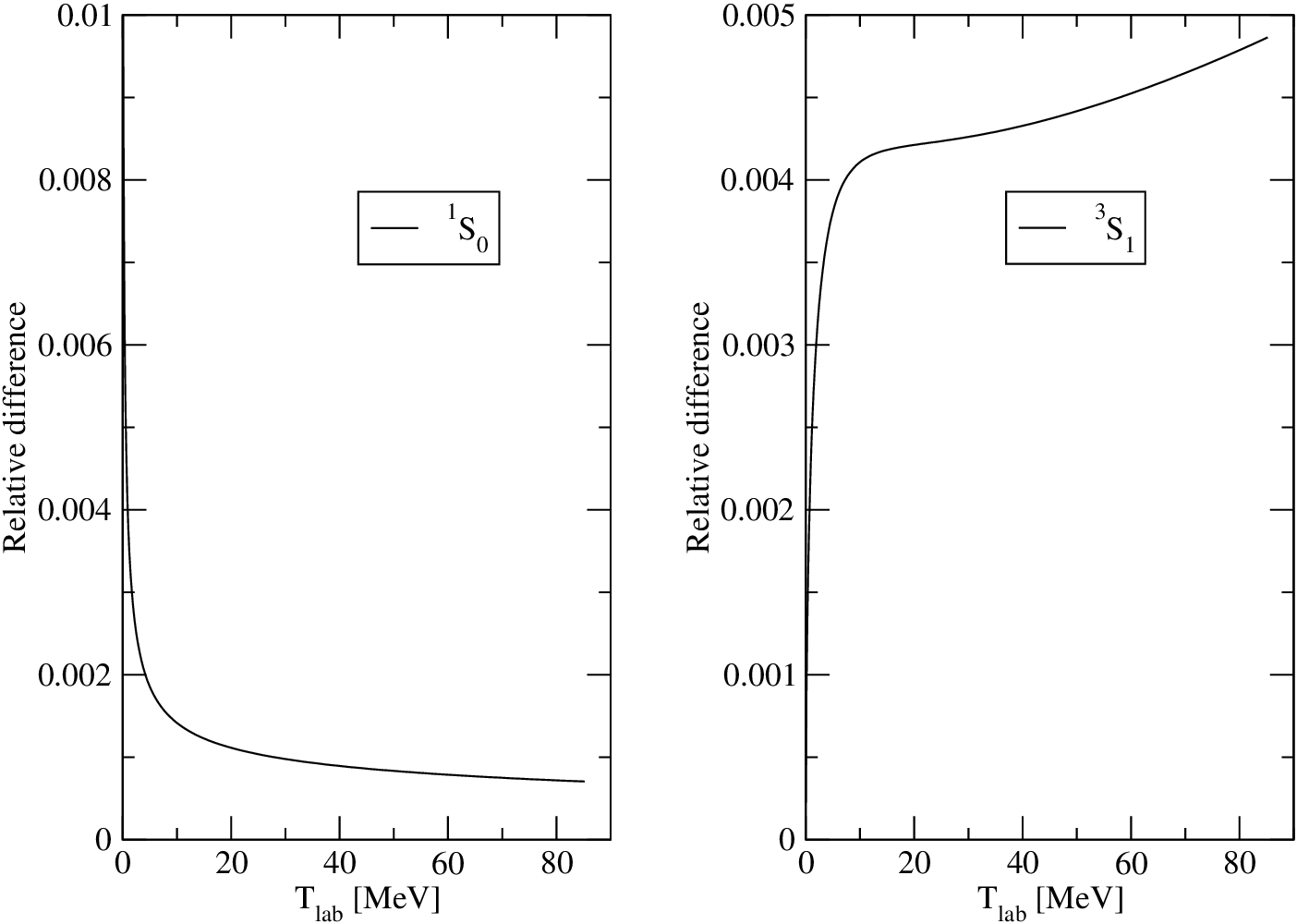}
\end{center}
\caption{ The relative difference of NN $\protect\delta(^1$S$_0)$
  and $\protect\delta(^3$S$_1)$ phase shift between two
  renormalization methods as a function of the laboratory kinetic
  energy $T_{\mathrm{lab}} \leq 80$ MeV. Here $\Lambda=50$ GeV is
  used, and 100 Gauss-Legendre quadrature points are chosen for the
  solution of the LS equation.}
\label{fig-fig6}
\end{figure}

For the same number of mesh points, converged phase shifts calculated by the two different
methods differ by 0.1--1.3\% depending on the energy, as is shown in
Fig.~\ref{fig-fig6}. 
The mixing parameter for the triplet channel is more sensitive to
numerical errors, and in that quantity there is a 1--2\% relative
difference between the two methods. However, we still claim that both 
methods are equivalent.

Frederico \textit{et al.} also follow the steps around
Eqs.~(\ref{eq:3.11}) and (\ref{eq:3.12}), although there the
difference of t-matrices is constructed between a given c.m. energy
$E$ of the NN system and a large
negative energy $-\frac{\mu^2}{M}$~\cite{Ti99}. (See also
Refs.~\cite{Keister:2005,Lin:2007}, where similar ``first-resolvent''
methods are employed.) Taking $\mu=0$ in Eq.~(8) of Ref.~\cite{Ti99}
yields our Eq.~(\ref{eq:3.15}). The difference between that work and
what we have done here is that in Ref.~\cite{Ti99} the Born
approximation $t=V$ with $V$ given by Eq.~(\ref{eq:opep}) is used to
determine $t(-\frac{\mu^2}{M})$. The final results are then
independent of the ``subtraction point'' $\mu$, provided that $\mu$ is
large enough. This procedure leads to numerical results for phase
shifts which are similar to ours. But it relies on the validity of the
Born approximation at large negative energies. Since the tensor
potential that operates in the triplet channel $\sim \frac{1}{r^3}$ at
short distances the Born approximation is not actually valid at any
energy~\cite{GW,ZK}.

\begin{figure}[tbp]
\begin{center}
\includegraphics[width=10cm]{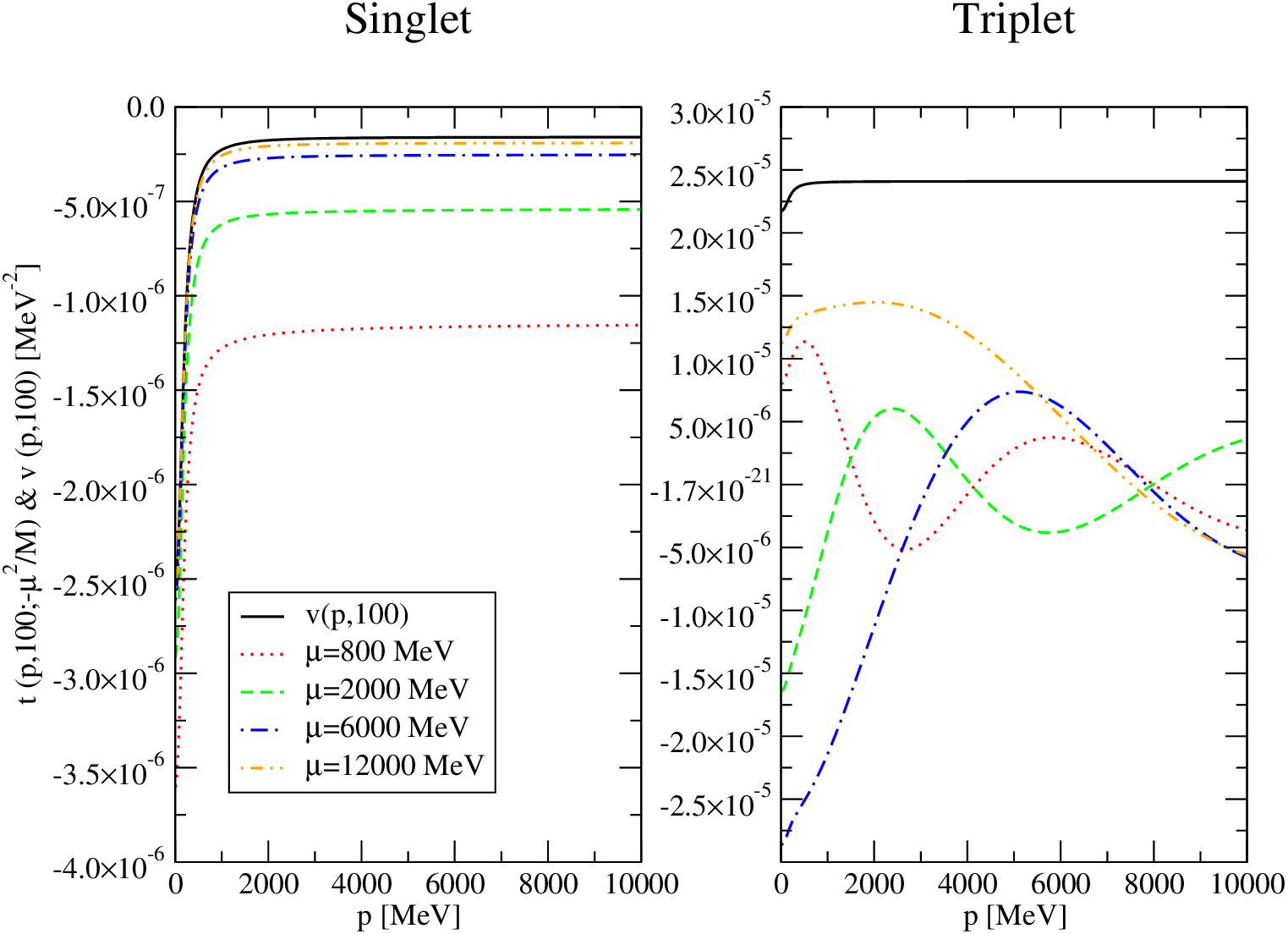}
\end{center}
\caption{The t-matrix elements $t^s(p,100;-\frac{\mu^2}{M})$ (left) and
$t^t_{00}(p,100;-\frac{\mu^2}{M})$ (right)
as functions of the half-shell momentum $p$ compared to 
$v^s(p,100)$ and $v^t_{00}(p,100)$. The parameter $\mu$
ranges from 0.8--12~GeV.}
\label{fig-fig13}
\end{figure}

\begin{figure}[tbp]
\begin{center}
\includegraphics[width=10cm]{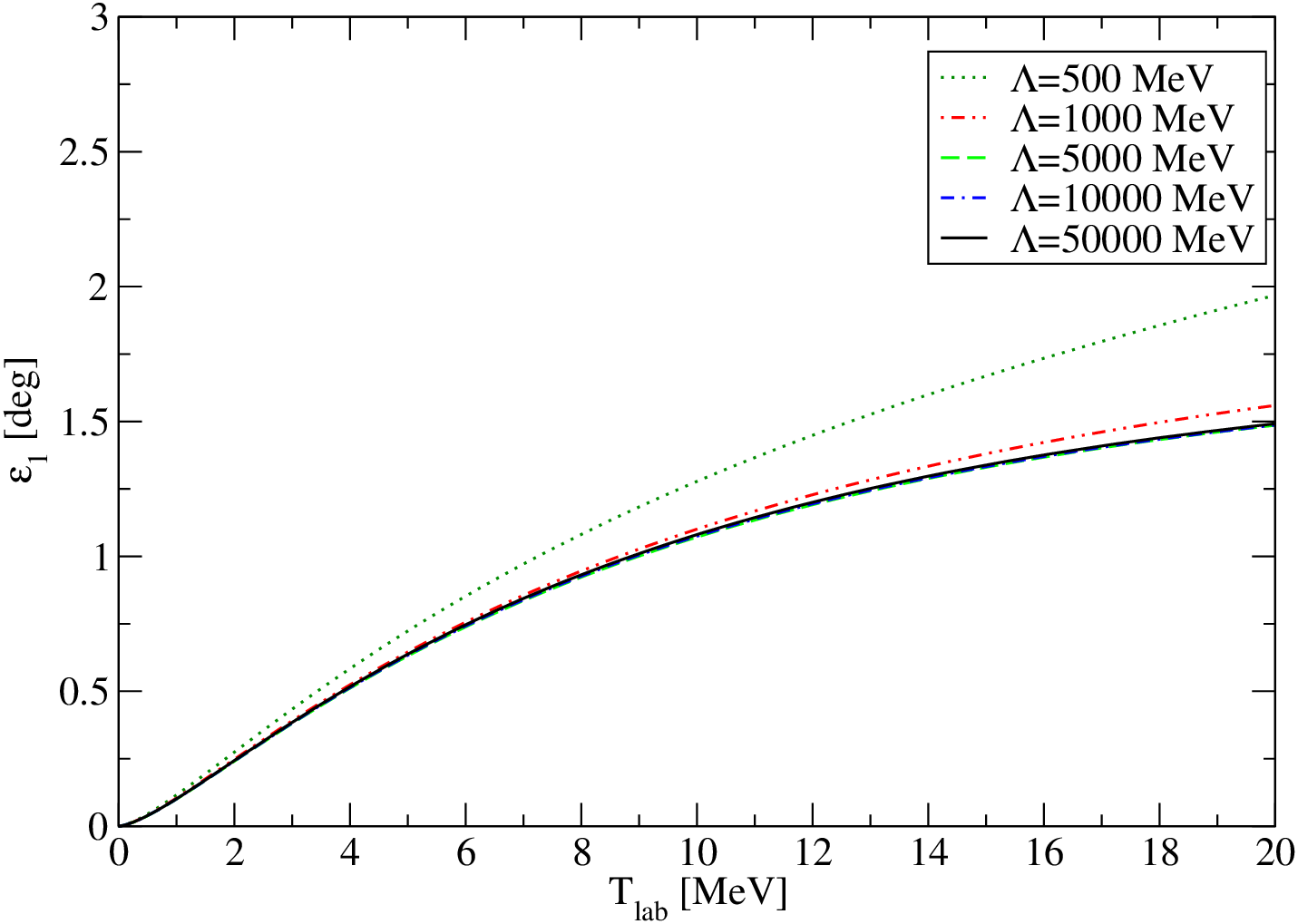}
\end{center}
\caption{The mixing parameter $\epsilon_1$ obtained with the subtraction method
as a function of the laboratory kinetic
energy $T_{\mathrm{lab}}\leq 20$~MeV and for cutoff parameters 
$\Lambda $ ranging from 0.5--50~GeV. 
}
\label{fig-fig14}
\end{figure}

We can see this failure of the Born approximation in the triplet
channel by examining the off-shell t-matrix at a number of negative
energies $E=-\frac{\mu^2}{M}$ and comparing the results to $V$
itself. Fig.~\ref{fig-fig13} shows that the behavior of
$t_t(p,p';-\frac{\mu^2}{M})$ at fixed $p'$ is completely different
from the behavior of $V$. (In contrast, in the singlet channel the
Born approximation appears to work quite well if $\mu$ is large
enough.)  This defect in Frederico {\it et al.}'s argument is manifested
in the fact that $\mu$ dependence in physical quantities disappears
only slowly as $\mu \rightarrow \infty$. The mixing parameter
$\epsilon_1$, which is particularly sensitive to the tensor potential,
is one significant example of this. Our subtraction method gives quite
good convergence of $\epsilon_1$ with respect to cutoff, as shown in
Fig.~\ref{fig-fig14}, and the convergence of other physical quantities
with respect to $\Lambda$ as $\Lambda \rightarrow \infty$ is quite
rapid, see Tables \ref{table-3}--\ref{table-1} (c.f. Table 1 of
Ref.~\cite{Ti99}).
This presumably results from the fact that our subtractive
renormalization of the LS equation is based only on Hermiticity of the
potential and momentum-independence of the contact interaction
$C$. Under these assumptions all the results of this section and
Sec.~\ref{sec-bound} below are determined by the form of one-pion
exchange and the value of the NN scattering lengths that are supplied
in Eqs.~(\ref{eq:3.4}) and (\ref{eq:3.5}).

\section{Subtractive Renormalization in Bound-State Calculations}
\label{sec-bound}

The neutron and proton form a bound state in the
$^{3}$S$_{1}$-$^{3}$D$ _{1}$ partial-wave state: the deuteron.  In
this section we review the methods by which the deuteron binding
energy is deduced from the NN t-matrix, and employ the t-matrix
obtained via subtractive renormalization in the previous section to
obtain NN binding energies and wave functions.  

\begin{figure}[tbp]
\begin{center}
\includegraphics[width=10cm]{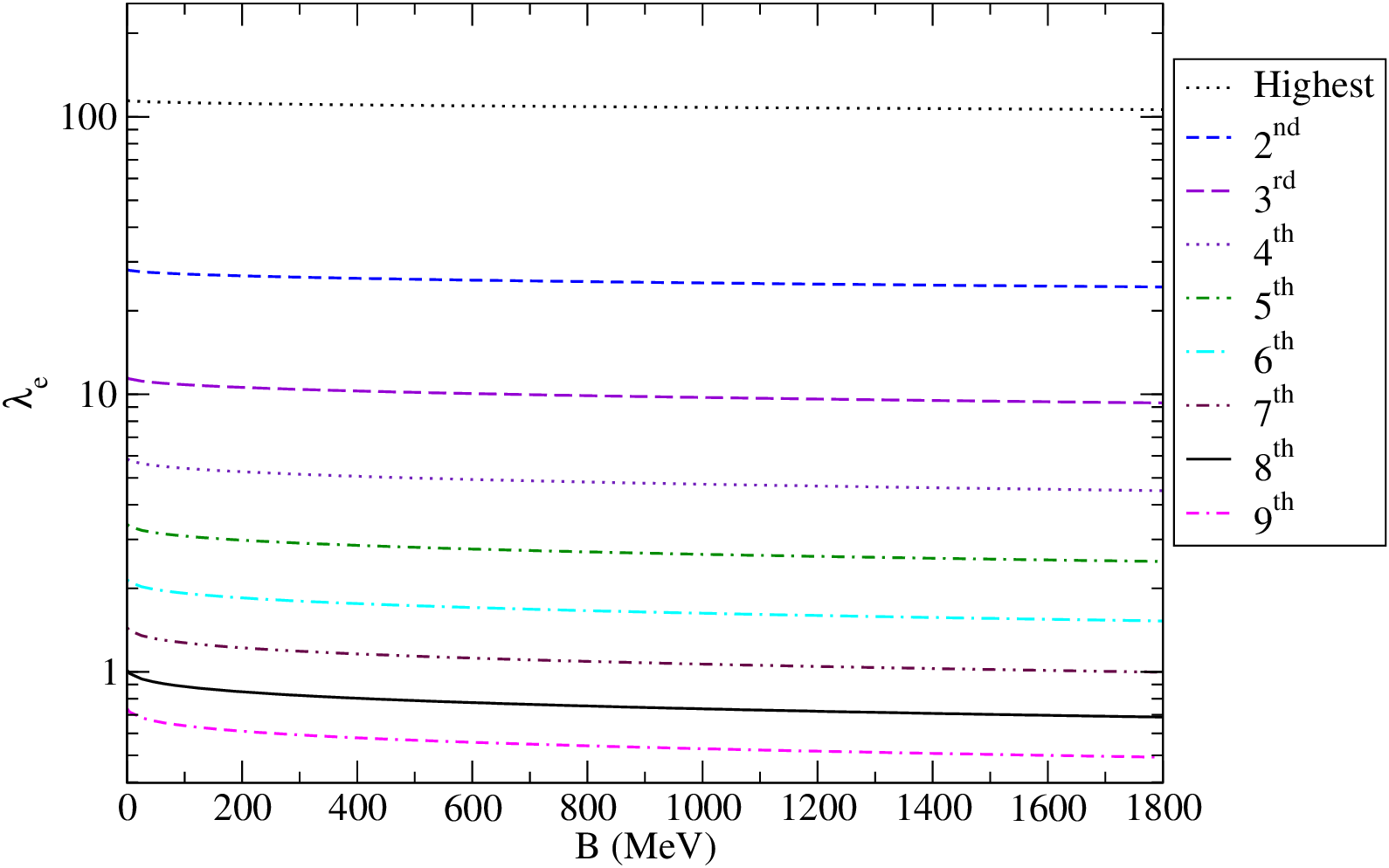}
\end{center}
\caption{(Color online) The eigenvalues $\protect\lambda_e$ obtained from
  Eq.~(\ref{eq:4.2}) as a function of the binding energy $B$. Note that
the vertical axis is on a log scale. }
\label{fig-fig7}
\end{figure}

Let us consider the Schr\"{o}dinger equation 
\begin{equation}
H\psi =E\psi =-B\psi   \label{eq:4.1}
\end{equation}
with $H=H_{0}+V$, $H_{0}=\frac{p^{2}}{M}$ the free Hamiltonian, $V$
the previously renormalized potential, and $B$ the unknown
binding energy. By rearranging and partial-wave expanding 
Eq.~(\ref{eq:4.1}) and defining: 
\begin{equation}
\langle p|\Gamma_l \rangle =(-B-\frac{p^{2}}{M})\langle p|\psi_l \rangle
=\langle p|(-B-H_{0})|\psi_l \rangle \equiv \langle p|g^{-1}(-B)|\psi_l
\rangle ,
\end{equation}
we obtain 
\begin{equation}
<p|\Gamma_l>=\frac{2}{\pi}
\int dp' \, p'^2 \, v_{ll'}(p,p^{\prime })\frac{1}{-B-\frac{p^{\prime 2}}{M}}
<p^{\prime }|\Gamma_{l'} >.  
\label{eq:4.2}
\end{equation}
The kernel $v_{ll'}(p,p^{\prime })\ (-B-\frac{p^{\prime 2}}{M})^{-1}$
of Eq.~(\ref{eq:4.2}) is the same as the kernel of the equation for
the t-matrix (\ref{eq:2.3}). We see from Eq.~(\ref{eq:4.2}) that the
kernel will have an eigenvalue $\lambda(E)=1$ when the energy is equal
to that at which a bound-state occurs, $E=-B$. Thus by varying $B$ and
solving the eigenvalue problem we can obtain the binding energy. If
the constant $C_T$ is fit to a scattering length of $a_{t}=5.43$~fm we
obtain $B=2.13$~MeV for $\Lambda=50000$ MeV. In Table~\ref{table-2} 
the values of $B \equiv B_{fit}$ for
different cutoff parameters are listed. 
We note that for small cutoff parameters $\Lambda$, the fitting procedure
leads to  a slightly larger value of $B_{fit}$.
In Fig.~\ref{fig-fig7} the trend of
eigenvalues $\lambda _{e}$ is shown 
as a function of $B_{fit}$ for $\Lambda=50000$ MeV, and we
see that the deuteron is actually the $8$th excited state of the
system for this cutoff. As $\Lambda \rightarrow \infty$
the number of bound states in the NN system increases
monotonically, but renormalization ensures that the shallowest one is
always at $B \approx 2$ MeV.

We now want to demonstrate that we can apply the approach discussed
in Sec.~3 to the bound-state problem. Multiplying Eq. (\ref{eq:3.15})
with $(E+B)$ we obtain 
\begin{equation}
(E+B) \ t(E) = t(0)\Big[(E+B)+\big(g(E)-g(0)\big) \ (E+B) \ t(E)\Big]
\label{eq:4.2b}
\end{equation}
When considering the bound state, i.e. $E\rightarrow -B$, the first
term in the brackets on the right-hand side vanishes. But, since, the
t-matrix has a pole at $E=-B$, we can define the
residue at the pole as $f(-B)=\underset{E\rightarrow -B}{\lim} (E+B)
\; t(E)$. Consequently, upon taking the limit, Eq.~(\ref{eq:4.2b}) becomes
\begin{equation}
f(-B)=t(0) \Big[g(-B)-g(0)\Big] f(-B) \equiv
t(0) \left[\frac{1}{-H_0-B}-\frac{1}{-H_0}\right]f(-B).
\label{eq:4.3}
\end{equation}
If this equation is written for the partial-wave form of the t-matrix then
the residue operator $f_{ll'}(-B)$ is related to the function
$\langle p|\Gamma_l \rangle$ of Eq.~(\ref{eq:4.2}) by
\begin{equation}
\langle p|\Gamma_l \rangle \langle \Gamma_{l'}|p' \rangle=\langle p|f_{ll'}(-B)|p' \rangle.
\end{equation}

\begin{figure}[tbp]
\begin{center}
\includegraphics[width=10cm]{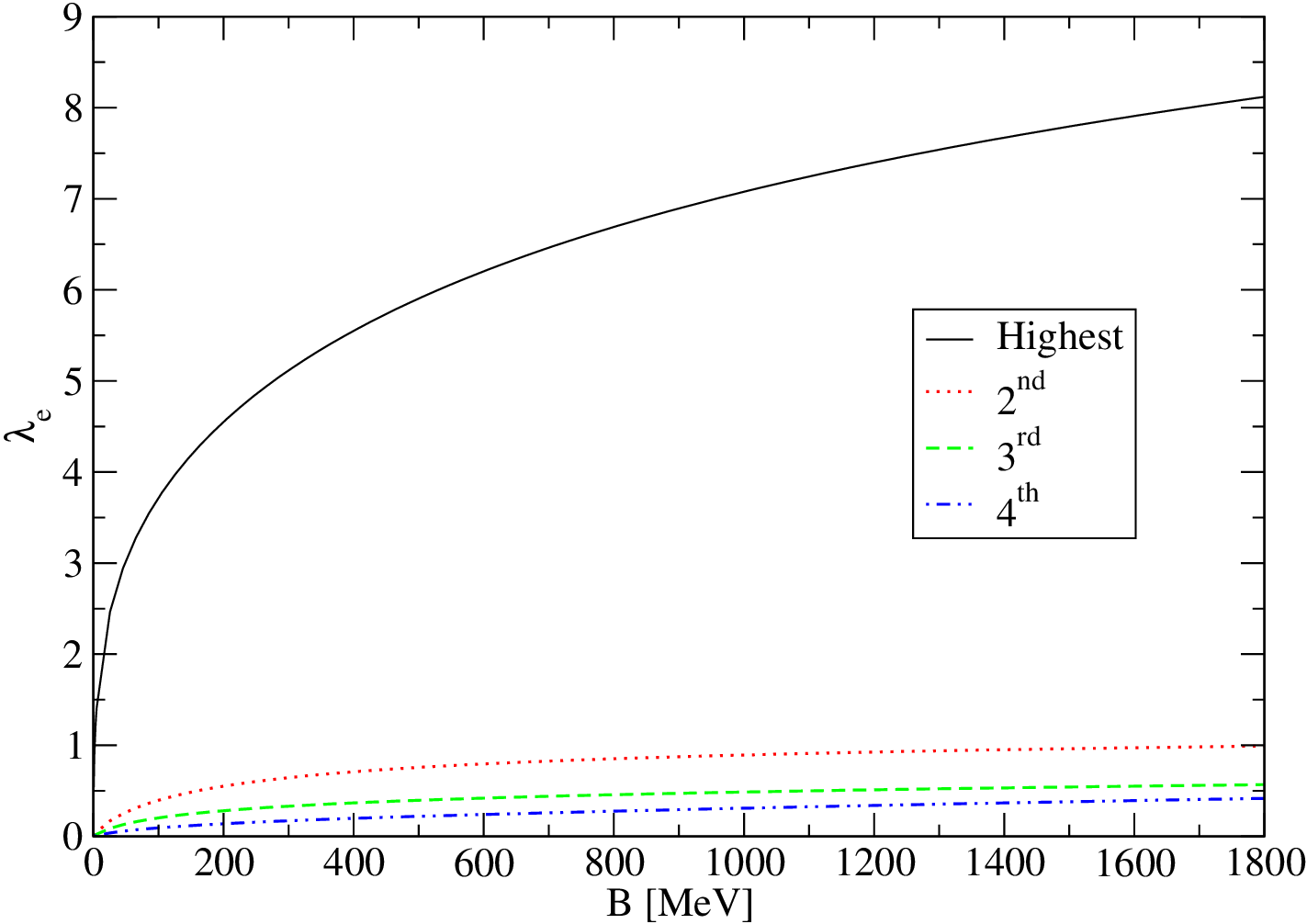}
\end{center}
\caption{(Color online)  The eigenvalues $\protect\lambda_e$ obtained from
  Eq.~(\ref{eq:4.3}) as a function of the binding energy $B$. }
\label{fig-fig8}
\end{figure}

We already calculated $t(0)$ in Section~3, thus the kernel of
Eq.~(\ref{eq:4.3}) can be evaluated directly. The energy $E=-B$ at
which it has an eigenvalue of 1 reveals the bound-state energy, and
the eigenvector is then the function $\langle p|\Gamma_l \rangle$ (up
to an overall constant).  Our numerical calculation gives a deuteron
binding energy $B\equiv B_{sub}=$2.11~MeV. This result is independent
of the cutoff to better than 0.5\% for $\Lambda > 500$ MeV (see Table~\ref{table-2}).  The relative
difference in the binding energy of the shallowest state obtained by
solving Eq.~(\ref{eq:4.2}) and Eq.~(\ref{eq:4.3}) is 0.94\% for
$\Lambda=50000$ MeV.  From this we conclude that the subtractive
renormalization works for the bound state as well (or better) than it
does for scattering observables.  Fig.~\ref{fig-fig8} shows the trend
of eigenvalues $\lambda_e$ of the kernel of Eq.~(\ref{eq:4.3}) as a
function of $B_{sub}$.  Our calculation gives a deuteron binding
energy different from the experimental value $B_{\rm exp}$=2.2246~MeV.
The reason for this difference is that we work only with the LO
$\chi$PT potential, and so the analytic continuation of the scattering
amplitude from the scattering to the bound-state regime is only
accurate up to corrections of order
$\frac{(p,m_\pi)^2}{\Lambda_\chi^2}$.

\begin{figure}[tbp]
\begin{center}
\includegraphics[width=10cm]{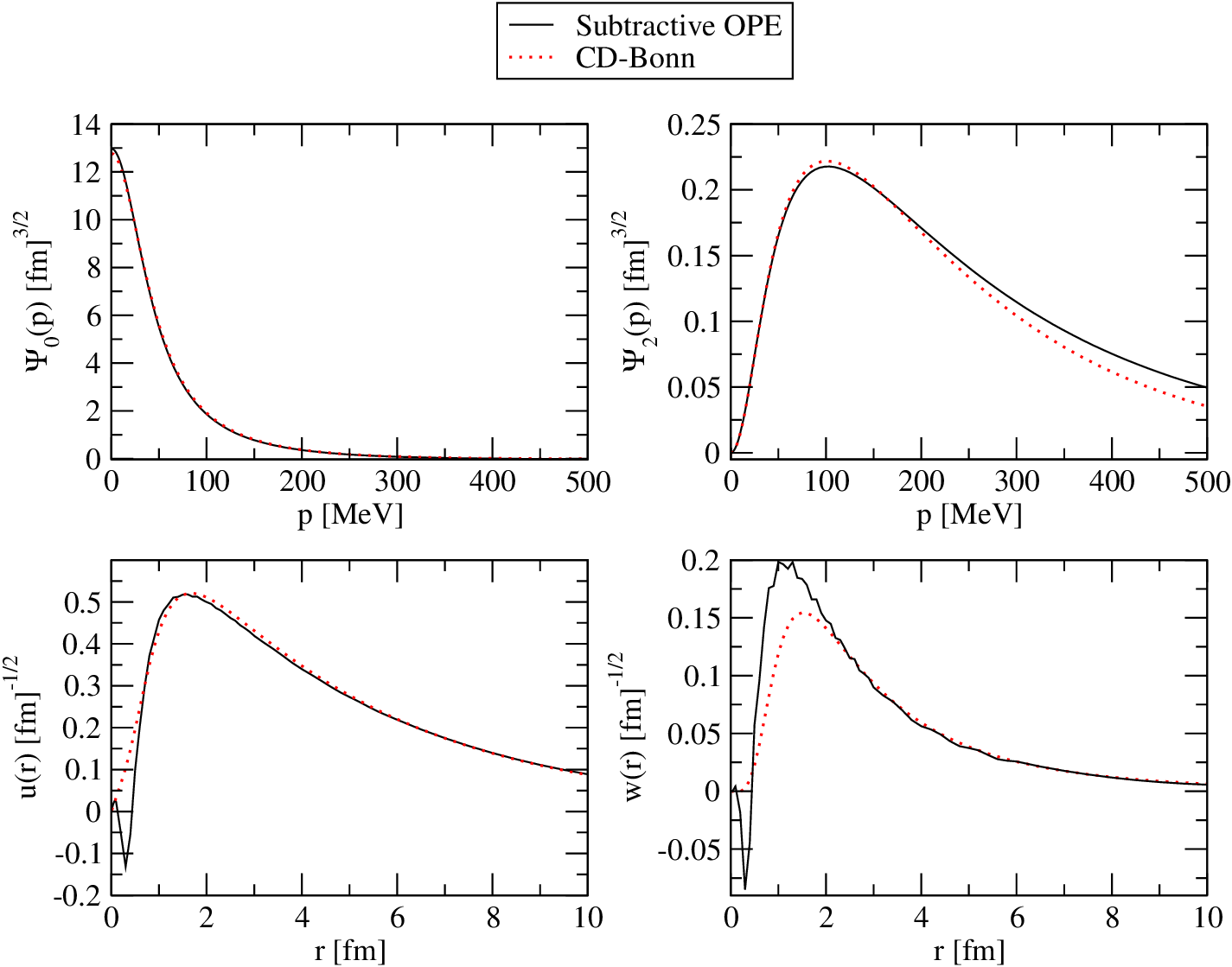}
\end{center}
\caption{(Color online) Upper panel shows the momentum-space wave functions of the shallowest $NN$
  bound state in the triplet channel 
as a function of p, where $\protect \psi_0$(p) is the $^3$S$_1$ wave
function
and
 $\protect\psi_2$(p) denotes the $^3$D$_1$ wave (solid lines). 
These wave functions
are obtained from the subtracted integral equation with $\Lambda=50$ GeV.
 Bottom panel shows the coordinate-space wave functions as functions of $r$, 
where $u(r)$ denotes the $^3$S$_1$ wave and
 $w(r)$ denotes the $^3$D$_1$ wave (solid lines).
The dotted lines indicate the corresponding wave functions obtained from
the CD-Bonn potential.}
\label{fig-fig9}
\end{figure}

In spite of this discrepancy with experiment we can still check that
our eigenenergy is the correct one for LO $\chi$PT with 
the experimental value of $a_t$ reproduced. To do this we 
use the S-wave effective-range expansion for the on-shell t-matrix 
$t_{00}^t $,
\begin{equation}
-Mt_{00}^t =\frac{1}{-\frac{1}{a_{t}}+\frac{1}{2}r_{t}
  p_0^{2}-ip_0}, \label{eq:4.5}
\end{equation}
with $r_t$ being the triplet effective range. This form of $t_{00}^t $ is valid
for on-shell momenta $p_0$ that obey $|p_0| < m_\pi/2$. The
pole in the t-matrix representing the shallowest bound state of the NN
system is inside
this radius of convergence. Thus we can
solve 
\begin{equation}
\frac{1}{a_{t}} - \frac{1}{2}r_t p_0^2+ip_0=0
\end{equation} 
for $p_0=i \gamma$ in order to determine its location,
and $B$ is then equal to $\gamma^2/M$. Using the
experimental value of $a_t$, and the value of $r_t$ extracted from our
phase shifts, we  obtain $B=2.07$ MeV. This value has a
relative difference of $\sim $~1\% with respect to the two previously
calculated values. This is consistent with the omission of terms of
$O(p_0^4)$ in the effective-range expansion form employed in Eq.~(\ref{eq:4.5}). 
We now turn to extraction of the deuteron wave function. In a
particular partial wave this is related to the above defined
function  $\langle p|\Gamma_l \rangle$ via
\begin{equation}
\psi_l(p)=-\frac{1}{B+\frac{p^{2}}{M}}\langle p|\Gamma_l \rangle .
\end{equation}
Since $\langle p|\Gamma_l \rangle$ is equal to the eigenvector, up to
an overall constant, it is now a simple matter to obtain $\psi_l(p)$. 
In the upper part of Fig.~\ref{fig-fig9}  the momentum-space S-wave and D-wave 
calculated from the above subtraction method are compared with 
those obtained from the CD-Bonn potential.
The overall constant is fixed by our
normalization condition
\begin{equation}
\int dp \, p^{2}[\psi _{0}^{2}(p)+\psi _{2}^{2}(p)]=1.
\end{equation}
Here $\psi _{0}(\psi _{2})$ is the $^{3}S_{1}(^{3}D_{1})$ wave
function. The agreement of the LO $\chi$PT momentum-space deuteron
wave function with that obtained from the CD-Bonn potential is
remarkable, especially in the S-wave.

Finally we obtain the coordinate space
wave functions by Fourier transformation
\begin{equation}
\frac{u_L(r)}{r}=\sqrt{\frac{2}{\pi }}\int dpp^{2}j_{L}(pr)\psi _{L}(p),
\end{equation}
where $j_{L}$\ is the spherical Bessel function, with $L=0$ for the
$^{3}$S$_{1}$\ wave and $L=2$ for the $^{3}$D$_{1}$ wave.  In
bottom part of Fig.~\ref{fig-fig9} the wave functions $u(r) \equiv u_0(r)$ and $w(r)
\equiv u_2(r)$ are shown as a function of $r$. We see that our
coordinate-space wave functions have a node near the origin, which is
due to the fact that the LO $\chi$PT potential has additional bound
states at larger values of $B$, i.e. the $B=2.1$ MeV state is not the
ground state. The short-distance behavior of the deuteron wave
function obtained from the LO $\chi$PT potential has been derived
analytically, and discussed in considerable detail, in
Ref.~\cite{PVRA05}. Our results for $u(r)$ and $w(r)$ at the
distances where the Fourier transform is numerically stable appear to
be in agreement with that work.

\begin{figure}[tbp]
\begin{center}
\includegraphics[width=10cm]{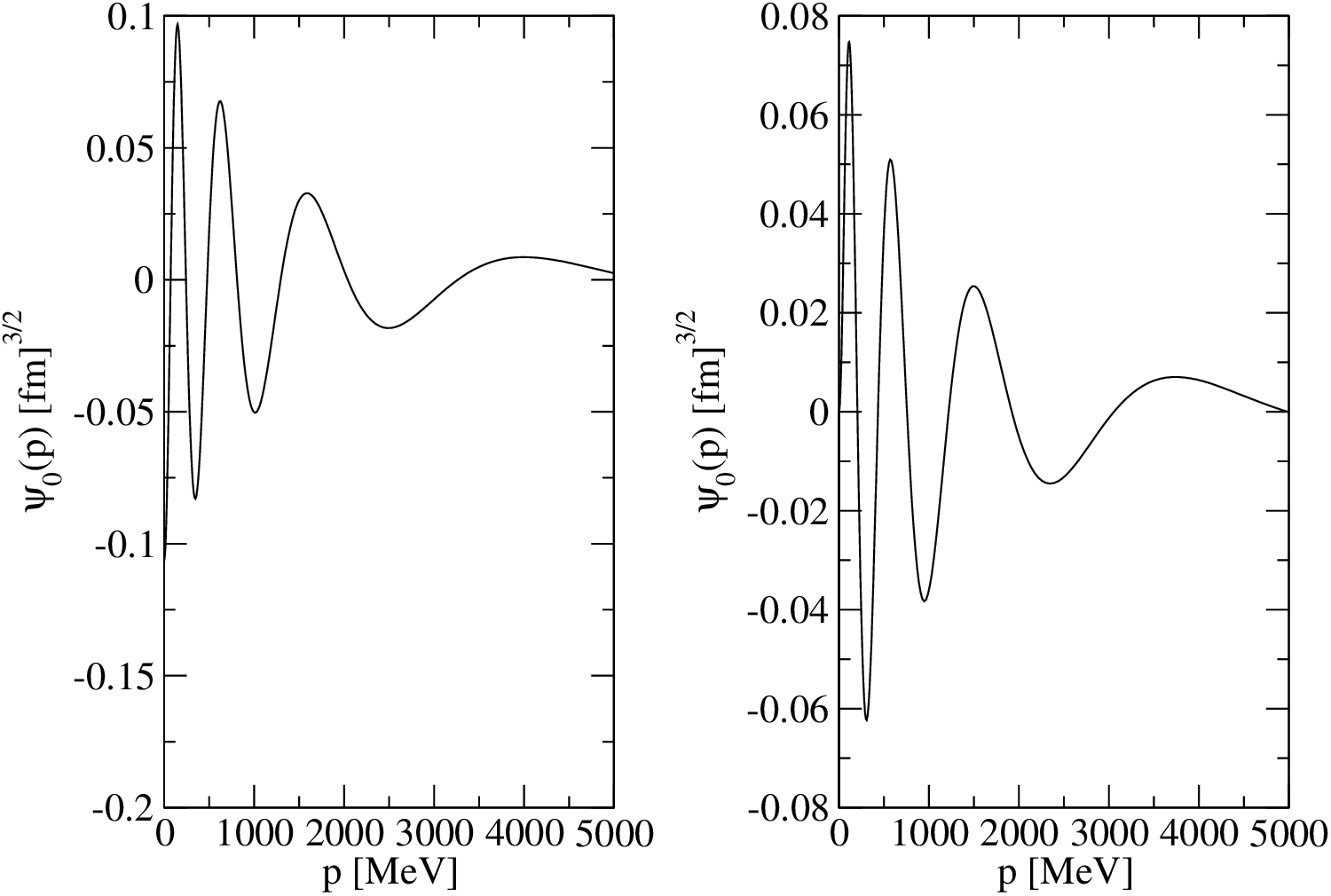}
\end{center}
\caption{Wave functions of the second-shallowest NN bound state in the
  triplet channel as functions of p, where
$\protect\psi_{0}$(p) denotes the $^3$S$_1$ wave and
  $\protect\psi_{2}$(p) 
denotes the $^3$D$_1$ wave. These wave functions are obtained from the
subtracted integral equation with $\Lambda=50$ GeV.}
\label{fig-fig11}
\end{figure}

In fact, Figs.~\ref{fig-fig7} and \ref{fig-fig8} show that  for this value of
the cutoff another eigenvalue crosses 1 at $B=1.76$ GeV. This is the
binding energy of the second-shallowest bound state for
$\Lambda=50000$ MeV; $B$ for this state
is slightly smaller at lower
cutoffs. The binding momentum of this state is about 1.2 GeV, and so
this state is outside the radius of convergence of $\chi$PT: the
theory's prediction of its existence is not reliable, and should be disregarded. For
completeness, we show the wave function corresponding to this binding
energy in Fig.~\ref{fig-fig11}. The momentum distribution confirms that
the prediction of this state by the low-momentum effective theory is
not reliable.

\section{Summary and Conclusions}

\label{sec-conclusion}

We have developed a subtractive renormalization of the S-wave
Lippmann-Schwinger equation for the leading-order NN potential of
$\chi$PT. Our results are equivalent to earlier work where the contact
interaction in NN S-waves was fitted to obtain a particular scattering
length. The advantage of our subtraction formulation is that all
reference to the unphysical, regularization- and
renormalization-dependent contact interaction $C$ disappears from the
scattering equation.  Instead of searching for the unknown constants
$C_S$ and $C_T$, we directly use the NN scattering lengths $a_s$ and
$a_t$ as input and thereby obtain low-energy phase shifts as well as
the spectrum and wave functions of all NN bound states which are
within the domain of validity of $\chi$PT. Finding $C_S$ and $C_T$ for
large cutoffs $\Lambda$ can be numerically challenging.  Subtractive
renormalization avoids this problem, and so allows us to compute $NN$
phase shifts from the LO $\chi$PT potential up to  $\Lambda=50$ GeV---a
factor of 10 larger than previously employed in momentum-space
calculations.

The subtractive technique hence yields phase shifts---and bound-state
information---based only on four assumptions:
\begin{enumerate}
\item The form of the long-range potential (in this case assumed to be
  one-pion exchange);
\item The validity of the Lippmann-Schwinger equation;
\item Hermiticity of the underlying NN potential;
\item Energy independence of the short-distance piece of the NN 
potential ---or at least approximate energy dependence over the range of
  energies considered.
\end{enumerate}
Our results therefore rest on fewer assumptions than were employed to
get NN phase shifts in the subtractive approach of
Ref.~\cite{Ti99}. In particular, we do not assume that the Born
approximation is valid for the unregulated NN potential (\ref{eq:opep}).

We have chosen to develop a subtraction based on using the NN
scattering lengths as input.  In principle experimental data at any
energy for which the effective theory is valid could be used to
determine $T(p_0,p_0;E)$ ($p_0^2=ME$), however, $E=0$ seems a natural
choice since it introduces the NN low-energy parameters directly into the LS
equations for the s-waves.  This freedom to choose the subtraction
point was exploited to develop Callan-Symanzik-type
renormalization-group equations for the NN system in
Refs.~\cite{Ti99,Ti05}.

For higher partial waves such as the NN P-waves (${}^3$P$_0$,
${}^3$P$_1$, $^{3}$P$_{2}$-$^{3}$F$_{2}$), we expect that similar
logic will allow us to obtain subtractive equations there
too. Furthermore, if we include the two-pion exchange
potential~\cite{Or96,Ka97}, which is the dominant correction to
one-pion exchange in the $\chi$PT expansion for the NN potential, then
we might hope that our results for phase shifts will be valid to
higher energies. The subtraction technique will still allow us to
renormalize the scattering equation for this long-range potential, as
long as the assumption that the constant $C$ is approximately energy
independent is still valid over the energy range considered (see also
Refs.~\cite{PVRA06B,Birse:2007}). The ability of subtractive
renormalization to facilitate higher-order calculations in this way
has already been demonstrated in the three-body problem in an
effective field theory with contact interactions alone~\cite{PP06,Platter:2006}.

\section*{Acknowledgments}
This work was funded by the US Department of Energy under grant
DE-FG02-93ER40756.  C.~E. and D.~P. thank the Department of Energy's
Institute for Nuclear Theory at the University of Washington for its
hospitality during the initial stages of the writing of this paper.
We also thank Enrique Ruiz Arriola and Manuel Pavon Valderrama for
illuminating explanations of their results and for comments on the
manuscript. We are also grateful to Lucas Platter for a careful
reading of the manuscript.



\clearpage
\begin{table}[tbp]
\begin{center}
\begin{tabular}{|c|c|c|}
\hline\hline
$\Lambda$ [MeV] & $r_s$ [fm] & $r_t$ [fm] \\ \hline
500 & $1.556$ & $1.582$ \\
1000 & $1.441$ & $1.614$ \\
5000 & $1.365$ & $1.596$ \\
10000 & $1.356$ & $1.595$ \\
50000 & $1.349$ & $1.594$ \\ \hline\hline
\end{tabular}
\vspace{0.3cm}
\end{center}
\caption{Effective range $r_s$(singlet) and $r_t$(triplet) 
obtained using the method of Section 2 for
various cutoffs $\Lambda$.}
\label{table-3}
\end{table}

\begin{table}[tbp]
\begin{center}
\begin{tabular}{|c|c|c|}
\hline\hline
$\Lambda$ [MeV] & $C_S$ [MeV$^{-2}$] & $C_T$ [MeV$^{-2}$] \\ \hline
500 & $-6.069\times 10^{-6}$ & $-2.307\times 10^{-6}$ \\ 
1000 & $-4.926\times 10^{-6}$ & $-1.838\times 10^{-5}$ \\
5000 & $-3.902\times 10^{-6}$ & $1.310\times 10^{-6}$ \\ 
10000 & $-3.753\times 10^{-6}$ & $2.040\times 10^{-5}$ \\ 
50000 & $-3.627\times 10^{-6}$ & $-4.201\times 10^{-8}$ \\ \hline\hline
\end{tabular}
\vspace{0.3cm}
\end{center}
\caption{The constants $C_S$ and $C_T$ as determined by fitting the
scattering lengths $a_s$ and $a_t$ for different cutoff parameters $\Lambda$.}
\label{table-1}
\end{table}

\begin{table}[tbp]
\begin{center}
\begin{tabular}{|c|c|c|}
\hline\hline
Mesh points & $\protect\delta (^{3}$S$_{1})$ Fitting & $\protect\delta (^{3}$S$_{1})$ Subtraction \\ \hline
20 & $139.2453$ & $104.6811$ \\ 
30 & $113.8653$ & $104.6649$ \\
40 & $107.9943$ & $104.6622$ \\ 
50 & $105.9794$ & $104.6618$ \\ 
100 & $104.2317$ & $104.6616$ \\ 
150 & $104.0749$ & $104.6616$ \\ \hline\hline
\end{tabular}
\vspace{0.3cm}
\end{center}
\caption{The convergence of the phase shift $\protect\delta (^{3}$S$_{1})$ at laboratory kinetic energy $T_{\mathrm{lab}}$=10 MeV as a function of the number of Gauss-Legendre quadratures for the fitting and subtraction methods. Note that the fitting method converges to 104.1 at about 200 mesh points, so the final results of these two methods differ by $\approx 0.5\%$ as per Fig.~5.}
\label{table-4}
\end{table}

\begin{table}[tbp]
\begin{center}
\begin{tabular}{|c|c|c|}
\hline\hline
$\Lambda$ [MeV] & $B_{fit}$ [MeV] & $B_{sub}$ [MeV] \\ \hline
500 & $2.10$ & $2.11$ \\
1000 & $2.12$ & $2.11$ \\
5000 & $2.11$ & $2.11$ \\
10000 & $2.11$ & $2.11$ \\
50000 & $2.13$ & $2.11$ \\ \hline\hline
\end{tabular}
\vspace{0.3cm}
\end{center}
\caption{The binding energy $B$ for different cutoff parameters $\Lambda$. 
The binding energy 
$B_{fit}$ is obtained from the fitting method described in Section~2, 
and the binding energy $B_{sub}$ is
 obtained from the subtractive method given in Section~3.}
\label{table-2}
\end{table}

\clearpage

\noindent 
\setlength{\unitlength}{1cm}

\par

\vspace{40mm}

\end{document}